\documentclass[reprint,superscriptaddress,nofootinbib,amsmath,amssymb,floatfix,float,aps,prd]{revtex4-2}
\usepackage{graphicx} 
\usepackage{dcolumn} 
\usepackage{bm} 
\usepackage[colorlinks=true, allcolors=blue]{hyperref}
\usepackage{cleveref}
\usepackage{xcolor}
\usepackage[normalem]{ulem}

\begin{document}

\title[]{Gauges for quadratic gravity: the extended transverse gauge and the energy-momentum tensor of the massive spin-2 field}

\author{Matheus F. S. Alves}
\email{matheus.s.alves@edu.ufes.br}
\affiliation{Departamento de Física, Núcleo de Astrofísica e Cosmologia (Cosmo-Ufes) \& PPGFis, Universidade Federal do Espírito Santo, Vitória, ES,  29075-910, Brazil.}
\author{L.G. Medeiros}
\email{leo.medeiros@ufrn.br}
\affiliation{Escola de Ci\^encia e Tecnologia, Universidade Federal do Rio Grande
do Norte, Campus Universit\'ario, s/n\textendash Lagoa Nova, CEP 59078-970,
Natal, Rio Grande do Norte, Brazil.}
\author{Davi C. Rodrigues}
 \email{davi.rodrigues@ufes.br}
\affiliation{Departamento de Física, Núcleo de Astrofísica e Cosmologia (Cosmo-Ufes) \& PPGFis, Universidade Federal do Espírito Santo, Vitória, ES,  29075-910, Brazil.}
\affiliation{Centro Brasileiro de Pesquisas Físicas (CBPF), R. Xavier Sigaud 150, 22290-180, Rio de Janeiro, RJ, Brazil}

\begin{abstract}
We study the 4D Einstein-Hilbert action extension based on the square of the curvature tensors. Analyses of gauge and perturbation modes are often done considering the Teyssandier gauge condition. Although this approach can be useful for other higher-order extensions of quadratic gravity, a generalized transverse (or Lorentz) gauge is clear and sufficient for the present case, as explained here. We provide a detailed analysis of the generalized transverse gauge condition, its residual gauge symmetry,  the physical modes, the induced energy-momentum tensor (EMT) of the massive spin-2 mode (which is gauge-dependent), and a comparison with the Teyssandier gauge. The derivation is valid for any EMT. We also compare the induced EMT in quadratic gravity with the massive spin-2 Fierz-Pauli EMT, which is different from the previous cases. The comparison is further developed by considering a spherical isothermal sphere, which works as an approximation to virialized spherical systems.
\end{abstract}

\maketitle

\section{Introduction}
This work focuses on quadratic gravity, a class of theories that extend General Relativity (GR) by including curvature-squared terms into the gravitational action \cite{Stelle:1977ry}. This theory is particularly attractive and was introduced in the context of quantum gravity, as it leads to a renormalizable framework \cite{Stelle:1976gc, Donoghue:2021cza,  Donoghue:2019clr}. However, the presence of higher-order derivatives may introduce Ostrogradsky instabilities, which manifest as ghost-like states upon quantization and pose significant challenges to unitarity \cite{Sbisa:2014pzo}. To address this issue, various approaches have been proposed, such as interpreting ghosts as unstable states or redefining the Hilbert space to preserve unitarity \cite{Salam:1978fd, Tomboulis:1977jk, 
Johnston:1987ue, Modesto:2015ozb, Anselmi2017, Donoghue:2019fcb, Salvio:2020axm}, see also \cite{Shapiro:2022ugk} and references therein for a recent review. Although a complete resolution remains open, new approaches  are being developed  (e.g., \cite{Camanho:2014apa, Edelstein:2021jyu}). Also, quadratic gravity and its extensions \cite{Asorey:1996hz, Salvio:2019wcp, Rachwal:2021bgb} continue to be used as effective theories to incorporate high-energy corrections to GR. In this work, we focus on classical aspects of such higher-derivative corrections, namely the gauge choices, propagating modes decomposition and in particular the dynamics of the massive spin-2 mode.

Quadratic gravity has been studied in a variety of contexts, including spherically symmetric solutions \cite{Lu2015a,Lu2015b, Antoniou:2024jku, Sajadi:2025nkm, Pravda:2024uyv, Kokkotas:2017zwt, Konoplya:2022iyn}, gravitational waves \cite{Alves:2022yea, Alves:2024gwi, Bogdanos2010, Holscher2019, Kim2019, Avijit2022}, and models of cosmic inflation \cite{Ivanov:2016hcm, Anselmi:2020lpp, DeFelice:2023psw}. In all these applications, consistent treatment of the theory requires careful gauge fixing due to the higher-order nature of the field equations. Although quadratic gravity preserves diffeomorphism invariance, as in GR, the presence of fourth-order derivatives makes gauge choices essential for simplifying the equations and isolating the physical degrees of freedom. 

The first works on quadratic gravity considered the transverse gauge (also known as the Lorentz or de Donder or harmonic gauge) \cite{Stelle:1977ry} (see also \cite{Salvio:2018crh} for further references), which is characterized by $\partial \bar h_{ab} = 0$, where $\bar h_{ab}$ is the trace-reversed metric perturbation about Minkowski. Currently, a particularly well-known gauge choice in quadratic gravity is the Teyssandier gauge \cite{Teyssandier1989}. In Sec.~\ref{sec:TeyssandierGauge} we review the latter gauge in detail. The basis of our main results are in Sec.~\ref{sec:ExtendedGauge} and are about a convenient extension of the transverse gauge, by using $\partial \bar h_{ab} = \partial_b \Theta$, where $\Theta$ is a scalar to be judiciously chosen. This type of gauge extension, although simple, is not commonly used in this context. The most useful choice for $\Theta$, the detailed explanation of this gauge, the residual gauge symmetries, the metric decomposition,  and the consequences for the massive spin-2-induced energy-momentum tensor are all detailed here. Most of the detailed analysis in the quadratic gravity literature is done in vacuum; here we pay considerable attention to the role of the matter environment. These developments are made in Secs.~\ref{sec:ExtendedGauge} \ref{sec:effectiveEMT}, \ref{sec:Comparisson}. The conclusions are presented in Sec.~\ref{sec:conclusions}. In the Appendix \ref{sec:app} we consider the spin-2 massive mode equation for a general $\Theta$. 

We stress that this work does not aim to directly resolve the issue of the ghost-like massive spin-2 mode. Our focus is on clarifying new possibilities and limitations related to gauge choices, which are not addressed in the existing literature.  Given the presence of the ghost, any direct phenomenological implications must be treated with caution. The use of smooth matter environments, as considered here, is intended primarily as a conceptual probe into the structure and consequences of quadratic gravity. We regard this endeavor as a meaningful step towards understanding and future applications of higher-derivative gravity.

We mostly use the same conventions as Wald \cite{Wald:1984rg}, in particular we use the signature $(-, +, +, +)$ and abstract index notation. That is, $A_{ab}$ refers to a covariant second-rank tensor (not the components). The components of $A_{ab}$ are denoted by $A_{\mu \nu}$ (with $\mu, \nu = 0, 1, 2, 3$), in a given coordinate system. The ordinary partial derivative in some coordinate system is $\partial_\mu$, while $\partial_a$ denotes a coordinate-independent covariant derivative that is compatible with Minkowski metric $\eta_{ab}$ in any coordinate system, that is, $\partial_a \eta_{bc} = 0$. Moreover, $\Box = \nabla_a \nabla^a$, $\Box^2 = \nabla_a \nabla^a \nabla_b \nabla^b$. For Minkowski spacetime, the previous derivatives can be replaced by $\partial_a$.

\section{The action and the perturbed field equations}

The action integral of quadratic gravity is given by \cite{Stelle:1976gc} 
\begin{equation} \label{eq:action}
S = \frac{1}{2 \kappa}\int d^4x \sqrt{-g} \left( R + \frac{1}{2} \gamma R^2
- \frac{1}{2} \alpha C^2 \right) + S_m \, .
\end{equation}
where $C^2 = C_{abcd}C^{abcd}$ is the contraction of the Weyl tensor, $S_{m}$
is the action of the matter fields, and $\kappa$, $\alpha$ and 
$\gamma$ are constants.

The field equations are (see also \cite{Alves:2022yea, Pravda:2024uyv, Sajadi:2025nkm})
\begin{gather}
R_{ab}-\frac{1}{2}g_{ab}R - 2 \alpha \left[ \nabla ^{c }\nabla ^{d}C_{a c b d }+
\frac{1}{2}R^{c d }C_{a c b d }\right] \label{field_eq_complete} \\
+\gamma \left[ R\left( R_{ab }-\frac{1}{4}Rg_{ab }\right) +g_{ab}\nabla
_{c}\nabla ^{c }R-\nabla _{a}\nabla _{b}R\right] =\kappa T_{ab}. \notag
\end{gather}
The trace of the field equation (\ref{field_eq_complete}) is 
\begin{equation}
3\gamma\Box R-R=\kappa T.  \label{eq:trace_field_equation}
\end{equation}
We expand the metric as 
\begin{equation}
g_{ab} = \eta_{ab} + h_{ab} 
\end{equation}
and introduce the trace-reversed metric perturbation, 
\begin{equation}
\bar h_{ab} = h_{ab} - \frac{1}{2} \eta_{ab} h \, .
\end{equation}

The first-order field equations are written as 
\begin{equation} \label{eq:QuadraticFieldEq}
G_{ab}^{(1)}+\gamma G_{ab}^{(\gamma )}+\alpha G_{ab}^{(\alpha )}=\kappa
T_{ab}\,,
\end{equation}
where 
\begin{equation}
G_{ab}^{(1)}=-\frac{1}{2}\left( \Box \bar{h}_{ab}+\eta _{ab}\partial
^{d}\partial ^{c}\bar{h}_{dc}-2\partial ^{c}\partial _{(a}\bar{h}
_{b)\,c}\right) \,, \label{eq:field_eq_1}
\end{equation}
\begin{eqnarray}
G_{ab}^{(\gamma )} &=&\eta _{ab}\Box \left( \frac{1}{2}\Box \bar{h}+\partial
^{d}\partial ^{c}\bar{h}_{dc}\right)  \notag \\
&&-\partial ^{d}\partial ^{c}\partial _{b}\partial _{a}\bar{h}_{cd}-\frac{1}{
2}\Box \partial _{b}\partial _{a}\bar{h}\,,  \label{eq:field_eq_2}
\end{eqnarray}
\begin{align}
G_{ab}^{(\alpha )}& =\frac{1}{2}\,\Box ^{2}\bar{h}_{ab}+\frac{1}{3}
\,\partial _{b}\partial _{a}\partial ^{d}\partial ^{c}\bar{h}_{cd}  \notag \\
& +\frac{1}{6}\,\partial _{b}\partial _{a}\Box \bar{h}-\,\partial
^{c}\partial _{(a}\Box \bar{h}_{b)c}  \notag \\
& +\frac{1}{6}\,\eta _{ab}\,\partial ^{d}\partial ^{c}\Box \bar{h}_{cd}-
\frac{1}{6}\,\eta _{ab}\,\Box ^{2}\bar{h}\,.  \label{eq:field_eq_3}
\end{align}

In the next sections, we will consider different gauge-fixing procedures. We start by revising the Teyssandier gauge in detail.

\section{The Teyssandier Gauge} \label{sec:TeyssandierGauge}

In this section we aim to do a detailed review on the Teyssandier gauge \cite{Teyssandier1989}, in the context of the action \eqref{eq:action}.  These results will be used to compare them with the extended transverse gauge, which is considered in the next section.

Considering the linearized field equation (\ref{eq:QuadraticFieldEq}), we define
    \begin{equation}
     \Gamma_{a} = (\alpha \Box - 1)\partial^{b} \bar{h}_{ba} + \frac{1}{3}(3\gamma - \alpha)\partial_{a} R,
    \end{equation}
    where the linearized Ricci scalar reads
    \begin{equation}
     R = \partial^{a} \partial^{b} \bar{h}_{ab} + \frac{1}{2}\Box \bar{h}. \label{R_e_h}
    \end{equation}
    Using $\Gamma_a$, Eq.~\eqref{eq:QuadraticFieldEq} can be written as
    \begin{eqnarray}
     (1 - \alpha \Box)\left( -\Box  \bar{h}_{ab} + \frac{1}{2} \eta_{ab} \Box  \bar{h} - \frac{1}{3} R \eta_{ab} \right)
     \nonumber \\
     + (\partial_{b} \Gamma_{a} + \partial_{a} \Gamma_{b}) = 2\kappa\left( T_{ab} - \frac{1}{3} \eta_{ab} T \right), \label{eq:fieldEqsGamma}
    \end{eqnarray}
    where we used Eq.~(\ref{eq:trace_field_equation}).
    The Teyssandier gauge \cite{Teyssandier1989} is defined by the condition
    \begin{equation}
     \Gamma_{a} = 0. \label{GT1}
    \end{equation}
    This condition is possible to be achieved since, under the infinitesimal diffeomorphism, 
    \begin{equation}
    {x'}^{a} = x^{a} + \xi^{a},
    \end{equation}
    $\Gamma_a$ transforms as
    \begin{equation}
     {\Gamma}_{a} \to \Gamma'_{a} = \Gamma_{a} - (1 - \alpha \Box)\Box \xi_{a}.
    \end{equation}

    Therefore, the condition $\Gamma'_{a} = 0$ can be realized by demanding that $\xi^a$ satisfies
    \begin{equation}
     (1 - \alpha \Box)\Box \xi_{a} = \Gamma_{a}.
    \end{equation}
The $\xi_a$ solution can be written as
\begin{equation}
  \xi_a = \zeta_a + \alpha \sigma_a \, ,
\end{equation}  
where the vector functions $\zeta_a$ and $\sigma_a$ obey the differential equations  
\begin{equation}
\square \zeta_a = \Gamma_a \quad \text{and} \quad (1 - \alpha \square)\sigma_a = \Gamma_a. \label{GT2}
\end{equation}  
It is possible to find $\zeta_a$ and $\sigma_a$ that satisfy above equations, since the former is a Poisson equation and the other an inhomogeneous Klein-Gordon equation. This concludes the demonstration that the gauge condition $\Gamma_a = 0$ is achievable. However, this condition is not sufficient to completely fix the gauge symmetry. Indeed, let
\begin{equation}
\tilde{x}^a = {x'}^a + \tilde{\xi}^a,
\end{equation}  
with $\tilde{\xi}_a = \tilde{\zeta}_a + \tilde{\sigma}_a$, where the new functions satisfy the homogeneous equations  
\begin{equation}
\square \tilde{\zeta}_a = 0 \quad \text{and} \quad (1 - \alpha \square)\tilde{\sigma}_a = 0.  \label{GTr}
\end{equation}  
Under these conditions, the Teyssandier gauge is preserved.

To solve Eq.~\eqref{eq:fieldEqsGamma} with $\Gamma_a = 0$ one writes $\bar{h}_{ab}$ as \cite{Teyssandier1989}
\begin{equation}
 \bar{h}_{ab} = \tilde{h}_{ab} + \tilde{\Psi}_{ab} - \eta_{ab} \Phi, \label{Metrica_decom}
\end{equation}
with
\begin{subequations}
\begin{align}
     & \Box \tilde{h}_{ab} = -2\kappa T_{ab}, \label{eq:GR} \\
     & \partial^a \tilde h_{ab} = 0 \, , \\
     & \Phi=-\gamma R.
\end{align}
\end{subequations}

From the trace equation, 
\begin{equation}
(\Box - m_{\Phi}^{2})\Phi = -\frac{\kappa}{3} T, \quad \text{with} \quad m_{\Phi}^{2} = \frac{1}{3\gamma}.
\end{equation}
From the above and Eq.~(\ref{eq:fieldEqsGamma}),
\begin{equation}
     (\alpha \Box - 1)\Psi_{ab} = 2\alpha \kappa\left( T_{ab} - \frac{1}{3} \eta_{ab} T \right)\, , \label{GTeqFinal}
\end{equation}
where
\begin{equation}
    \Psi_{ab}=\tilde{\Psi}_{ab}-\frac{1}{2} \eta_{ab}\tilde{\Psi}. \label{reversed_psi}
\end{equation}

An additional constraint on ${\Psi}_{ab}$ can be derived from Eq.~\eqref{R_e_h}, 
\begin{equation}
\partial^a \partial^b {\Psi}_{ab} = \square {\Psi}.  \label{vinculo_Psi}
\end{equation}

The field equations in the linear regime, therefore, can be written as
\begin{subequations}\label{eq:TeyssandierEqs}
\begin{align}
& \square \tilde{h}_{ab} = -2\kappa T_{ab},  \\
& (\square - m_{\Phi}^{2}) \Phi = -\frac{\kappa}{3} T,  \\
& (\Box - m^2_{\Psi}){\Psi}_{ab} = - 2\kappa T_{ab}^{\rm T} \, , \label{eq:PsiEquation} \\ 
& \partial^{a} \tilde{h}_{ab} = 0 \, , \\ 
& \partial^{a} \partial^{b} \Psi_{ab} = \Box \Psi \, , \label{eq:PsiConditionTey}
\end{align}
\end{subequations}
with
\begin{subequations}
\begin{align}
    & \bar{h}_{ab} = \tilde{h}_{ab} + \Psi_{ab} - \eta_{ab}\left( \Phi + \frac{1}{2}\Psi \right) \, , \label{eq:TeyssandierDecomp}\\
    & T^{\rm T}_{ab} =  - T_{ab} + \frac{1}{3} \eta_{ab} T \, . \label{eq:TeyssandierT}
\end{align}
\end{subequations}
$T^{\rm T}_{ab}$ is here called the Teyssandier-induced energy-momentum tensor for the massive field $\Psi$. 

For the full linear dynamics, there is still a piece of information that is not explicit in the equations above: the residual gauge. In vacuum $T_{ab} = T^{\rm T}_{ab} = 0 $ and it is possible to set $\tilde \zeta_a$ and $\tilde \sigma_a$ such that $\Psi =0$ and $\partial^a \Psi_{ab}=0$, which are standard conditions for a massive spin-2 mode in vacuum \cite{Fierz:1939ix}. However, it is clear from Eq.~\eqref{eq:PsiEquation} that for a general $T_{ab}$  these additional conditions are not possible.

\section{The extended transverse gauge}\label{sec:ExtendedGauge}

Since the field equations come from a (scalar) action that only depends on the metric, and since the trace reversed metric perturbation transforms as \cite{Wald:1984rg}
\begin{equation}  \label{eq:hGaugeTransf}
\delta \bar h_{ab} = \bar h'_{ab} - \bar h_{ab}= \partial_a \xi_b + \partial_b \xi_a - \eta_{ab}
\partial^c \xi_c \, ,
\end{equation}
under a diffeomorphism with $x^{\prime a} = x^a + \xi^a$, one can always choose $\xi^a$ such that
\begin{equation}
\partial^a \bar h_{ab} = \partial_b \Theta,  \label{eq:ThetaGauge_1}
\end{equation}
where $\Theta$ is some scalar. The transverse gauge (also called the Lorentz gauge) is a particular case with $\Theta = 0$. The condition \eqref{eq:ThetaGauge_1} is achievable since $\partial^a \bar h'_{ab} = \partial_b\Theta$ implies $\partial_b\Theta = \partial^a \bar h_{ab} + \Box \xi_b$. Therefore,  the vector field $\xi_a$ is chosen to satisfy the inhomogeneous wave equation $\Box \xi_b = \partial_b\Theta - \partial^a \bar h_{ab}$. We call this gauge condition, with arbitrary $\Theta$, extended transverse gauge.   

Since any new diffeomorphism with $\Box \tilde \xi^a = 0$ preserves the gauge \eqref{eq:ThetaGauge_1}, there is a residual gauge. This residual gauge symmetry cannot be used to fix the trace or other components of $\bar h_{ab}$, since $\bar h_{ab}$ is not a harmonic function (even assuming $T_{ab} = 0$). That is, from Eq.~\eqref{eq:hGaugeTransf}, one can use the residual gauge transformation $\tilde \xi_a$ to change the trace of $\bar h_{ab}$ as $\bar h' = \bar h - 2 \partial^a \tilde \xi_a$. But, since $\bar h$ is arbitrary, the divergence of a harmonic function cannot be used to eliminate $\bar h$. On the other hand, the residual gauge can be used to set $\bar h^{\mathrm{H}} = 0$ and $\bar h_{0a}^{\mathrm{H}} = 0$, where $\bar h_{ab}^{\mathrm{H}}$ is the harmonic part of $\bar h_{ab}$.

From Eq.~\eqref{eq:ThetaGauge_1},  without loss of generality, we write 
\begin{equation}
\bar{h}_{ab}=\xi _{ab}+\eta _{ab}\Theta \,,  \label{eq:h=XiTheta}
\end{equation}
where $\xi _{ab}$ is divergence-free: $\partial ^{a}\xi _{ab}=0$. With the
gauge and the decomposition above, the field equation \eqref{eq:QuadraticFieldEq} reads 
\begin{gather}
\partial _{a}\partial _{b}\Theta -\eta _{ab}\Box \Theta -\frac{1}{2}\Box \xi
_{ab} \label{eq:xiFieldEq1}\\
+\gamma \left( -3\partial _{a}\partial _{b}\Box \Theta -\frac{1}{2}\partial
_{a}\partial _{b}\Box \xi +3\eta _{ab}\Box ^{2}\Theta +\frac{1}{2}\eta
_{ab}\Box ^{2}\xi \right)  \notag \\
+\alpha \left( \frac{1}{6}\partial _{a}\partial _{b}\Box \xi +\frac{1}{2}
\Box ^{2}\xi _{ab}-\frac{1}{6}\eta _{ab}\Box ^{2}\xi \right) =\kappa
T_{ab}\,, \notag
\end{gather}
with $\xi \equiv \eta^{ab} \xi_{ab}$. The trace of the above equation is a particular case of Eq.~\eqref{eq:trace_field_equation} and it reads
\begin{equation}
-\Box (1-3\gamma \Box )\frac{1}{2}\left( 6\Theta +\xi \right) =\kappa T,
\label{eq:xiFieldEq1_trace}
\end{equation}
Setting the $\Phi $ field as
\begin{equation}
\Phi =-\frac{\gamma }{2}\Box (6\Theta +\xi )\,,  \label{eq:PhiDef}
\end{equation}
which, up to first order, is equivalent to $\Phi =-\gamma R$, we rewrite the Eq.~(\ref{eq:xiFieldEq1_trace}) as 
\begin{equation}
\left( \Box -m_{\Phi }^{2}\right) \Phi =-\frac{\kappa }{3}T\,,
\end{equation}
with $m_{\Phi }^{2}=1/(3\gamma )$. As expected, this is the same mass found in the Teyssandier gauge.

Since $\Theta$ is arbitrary (it can, for instance, be set to zero), one might be tempted to choose $\Theta = - \xi / 6$, which would imply $\Phi = 0$, suggesting that $\Phi$ is pure gauge. However, since $\Phi = -\gamma R$, this assumption is not valid. Similarly, choosing $\Theta = - \xi / 4$ would imply $\bar h = 0$, which is not a condition attainable through a gauge transformation either. More generally, it is not valid to consider $\Theta$ as a function of $\xi_{ab}$ for the purpose of gauge fixing. To impose the gauge condition~\eqref{eq:ThetaGauge_1}, one starts with an arbitrary $\bar h_{ab}$ and $\Theta$, and then determines a coordinate transformation (obtained by solving an inhomogeneous wave equation) that leads to a new $\bar h_{ab}$ satisfying Eq.~\eqref{eq:ThetaGauge_1}. The resulting $\bar h_{ab}$ can then be written in terms of $\xi_{ab}$ and $\Theta$. Since $\xi_{ab}$ is only defined \textit{after} $\Theta$, the demonstration above cannot be used to consider $\Theta$ as a function of $\xi^a$, and the counterexamples just shown assure that demanding such dependence can lead to wrong statements. 

The next step is to derive the tensor-mode equations. First we note that the $\gamma$-dependent term in Eq.~\eqref{eq:xiFieldEq1} can be written as
\begin{align}
\gamma \Box \left[ (-\partial _{a}\partial _{b}+\eta _{ab}\Box )  \left(
3\Theta +  \frac{1}{2}\xi \right) \right]  & =  \\[.2cm] 
& \!\!\!\!\!\!\!\!\!\!\!\!\! = (\partial _{a}\partial _{b}-\eta_{ab}\Box )\Phi \,. \nonumber
\end{align}
Therefore, Eq.~(\ref{eq:xiFieldEq1}) reads
\begin{gather}
(\partial _{a}\partial _{b}-\eta _{ab}\Box )\left( \Theta +\Phi +\frac{
\alpha }{6}\Box \xi \right) + \nonumber \\ +\frac{1}{2}\left( -1+\alpha \Box \right) \Box \xi _{ab}=\kappa T_{ab}\,.
\end{gather}
As expected, both sides of the above equation are divergence-free.
 
We now decompose the divergence-free part of $\bar h_{ab}$ into
\begin{equation}
\xi_{ab}=\tilde{h}_{ab}+\Psi_{ab}\,,
\end{equation}
where $\tilde h_{ab}$, by definition (and just like in the previous section), satisfies $\partial^a \tilde h_{ab}=0$ and the field equation of a massless spin-2 field
\begin{equation}
\Box \tilde{h}_{ab}=-2\kappa T_{ab}\,.  \label{eq:htildeDef}
\end{equation}
The solution for $\tilde h_{ab}$ can be written as the sum of a harmonic term (that satisfies the homogeneous differential equation above) plus a non-harmonic one. Without loss of generality, we assume that the harmonic part of $\xi_{ab}$ is entirely contained in\footnote{That is, $\Psi_{ab}$ belongs to the linear space of non-harmonic solutions, in particular $\Box \Psi_{ab} =0  \implies \Psi_{ab} = 0$, and linear combinations of $\Psi_{ab}$ belong to the same subspace.} $\tilde h_{ab}$.

For $T_{ab}=0$, $\tilde{h}_{ab}$ is harmonic, thus one can use the residual gauge to change $\tilde{h}_{ab}$ such that it satisfies $\tilde{h}=0$ and $\tilde{h}_{0a}=0$, as usual in the TT gauge of GR.

For arbitrary $T_{ab}$, 
\begin{gather}
-\frac{1}{2}\Box \Psi _{ab}+\frac{\alpha }{2}\Box ^{2}(\tilde{h}_{ab}+\Psi
_{ab})  \notag \\
+(\partial _{a}\partial _{b}-\eta _{ab}\Box )\left( \Theta +\Phi +\frac{
\alpha }{6}\Box (\tilde{h}+\Psi )\right) =0\,. \label{eq:boxPsi_general0}
\end{gather}

Using that the trace of the above equation reads
\begin{equation}
\Box \Psi =-6\Box (\Theta +\Phi )\,,  \label{eq:boxPsi_general}
\end{equation}
the field equation for $\Psi _{ab}$ is 
\begin{gather}
\left( \Box -\frac{1}{\alpha }\right) \Psi _{ab}=2\kappa T_{ab}+ \label{eq:general_extended_psi} \\
+\left( \frac{\partial _{a}\partial _{b}}{\Box }-\eta _{ab}\right) \left[ 
\frac{2\kappa }{3}T+2\left( \Box -\frac{1}{\alpha }\right) \Theta +2\left(
\Box -\frac{1}{\alpha }\right) \Phi \right] .  \notag
\end{gather}

A natural gauge condition is $\Theta = -\Phi$, which implies
\begin{equation}
\bar{h}_{ab} = \tilde{h}_{ab} + \Psi_{ab} - \eta_{ab}\Phi \, . \label{eq:ExtendedDecomp}
\end{equation}
This metric decomposition should be contrasted to that of the Teyssandier gauge \eqref{eq:TeyssandierDecomp}. We stress that within this gauge approach, we cannot use fix $\Theta$ to make a new $\Psi$ term to appear, since that would require $\Theta$ to be a function of $\xi_{ab}$. The above decomposition may seem to be equal to that of Eq.~\eqref{Metrica_decom}, but $\tilde \Psi_{ab}$ does not satisfy the same conditions of $\Psi_{ab}$ in this gauge.\footnote{We add that, although we use the symbol $\Psi_{ab}$ for both gauges, they are not exactly the same field, since they are defined from different conditions.} In conclusion, the metric decomposition in Eq.~\eqref{eq:ExtendedDecomp} is sensible in the extended transverse gauge context, but not in the Teyssandier case.

In this gauge with $\Theta = - \Phi$, Eq.~\eqref{eq:general_extended_psi} becomes remarkably simple,
\begin{equation}
\left( \Box - \frac{1}{\alpha} \right) \Box \Psi_{ab} = 2 \kappa \left[ \Box
T_{ab} + \frac{1}{3}(\partial_a \partial_b - \eta_{ab} \Box) T \right] \, . 
\label{eq:psi_t_psi}
\end{equation}
Although the above $\Theta$ choice seems to be the most useful in general, in Appendix~\ref{sec:app} we explore other $\Theta$ possibilities.

To understand the above equation, we start from the simplest case, which is given by vacuum $T_{ab} = 0$ or, with more generality, $\Box T_{ab} = 0$. Since $\Psi_{ab}$ belongs to the non-harmonic part of $\xi_{ab}$, the single relevant solution is a massive one with $m^2_\Psi =  \alpha^{-1}$,
\begin{eqnarray}
  \left( \Box - m^2_\Psi \right) \Psi_{ab} = 0 \, .   \label{eq:homogeneousPsi}  
\end{eqnarray}

In general,  Eq.~\eqref{eq:psi_t_psi} can be written as,
\begin{equation}
\left( \Box - m^2_\Psi \right) \Psi_{ab} = - 2 \kappa T^{\rm E}_{ab} ,
\end{equation}
with 
\begin{equation}
T^{\rm E}_{ab} \equiv \begin{cases}
    0  \mbox{ , if $\Box T_{ab} = 0$} \\[.4cm]
    -T_{ab} + \frac{1}{3}\left( \eta_{ab} - \frac{\partial_a \partial_b}{\Box} \right) T  \mbox{ , if $\Box T_{ab} \not= 0$}
\end{cases} . \label{eq:def_psi_t}
\end{equation}

Two direct and important properties of the energy-momentum tensor for the massive tensor mode within the extended-transverse gauge, denoted by $T^{\rm E}_{ab}$, are
\begin{subequations}
\begin{align}
    & T^{\rm E} = 0 \, , \label{eq:psiTrace}\\[.3cm]
    & \partial^a T^{\rm E}_{ab} = 0 \, .
\end{align}
\end{subequations}
We note that in this gauge with $\Theta = -\Phi$, Eq.~\eqref{eq:boxPsi_general} implies $\Box \Psi = 0$, thus implying $\Psi = 0$; since $\Psi_{ab}$ has no harmonic part from its definition. Hence, $T^{\rm E}_{ab}$ needs to be traceless, and thus compatible with $\Psi = 0$, as imposed by Eq.~\eqref{eq:psiTrace}. One sees that $\Psi_{ab}$ indeed satisfies the constraints of a massive spin-2 field (i.e., both its divergent and trace are zero, implying 5 degrees of freedom for the massive tensor mode alone) \cite{Fierz:1939ix, VanNieuwenhuizen:1973qf}.

Considering the expression in Eq.~\eqref{eq:def_psi_t}, we stress that $\partial_a$, likewise in  Ref.~\cite{Wald:1984rg}, denotes a derivative compatible with the Minkowski metric $\eta_{ab}$ in any coordinate system. It coincides with an ordinary partial derivative ($\partial_\mu$) only in the canonical coordinates of Minkowski spacetime, that is , if $(\eta_{\mu \nu}) = \mbox{diag} (-1, 1, 1, 1)$.

The complete set of field equations and constraints, in this gauge, is therefore
\begin{subequations} \label{eq:ExtendedEqs}
\begin{align}
  & \Box \tilde{h}_{ab}=-2\kappa T_{ab}\,, \\[.2cm]  
  & \left( \Box -m_{\Phi }^{2}\right) \Phi =-\frac{\kappa }{3}T\,, \label{eq:ETscalar}\\[.2cm]   
  & \left( \Box - m^2_\Psi \right) \Psi_{ab} = - 2 \kappa T^{\rm E}_{ab} \, , \label{eq:ETGPsi}\\[.2cm]
  & \partial^a \tilde h_{ab} = \partial^a \Psi_{ab} = 0 \, , \label{eq:ETGcond} \\[.2cm]
  & \Psi = 0 \, \label{eq:Psi0}.
\end{align}
\end{subequations}

The massive tensorial mode has an effective energy-momentum tensor that is similar to the one directly derived from the Fierz-Pauli massive spin-2 equation \cite{Fierz:1939ix, VanNieuwenhuizen:1973qf} (see also \cite{Finn:2001qi, Poddar:2021yjd}), namely, adapting to the present notation,
\begin{equation} \label{eq:TpsiFP}
    T^{\rm FP}_{ab} = T_{ab} + \frac{\partial_a \partial_b - \eta_{ab} \Box}{3 m^2_\Psi} T \, .
\end{equation}
Apart from the already expected inverted sign, $T^{\rm FP}_{ab}$ depends on $m_\Psi$, while $T^{\rm E}_{ab}$ has a dependence on $\Box^{-1}$. Since $T^{\rm E}_{ab}$ is an effective energy-momentum tensor that comes from a high-order derivative theory, the dependence on $\Box^{-1}$ is not surprising.
Clearly $\partial^a T^{\rm FP}_{ab} = 0$, just like $T^{\rm E}_{ab}$. With respect to the trace, a difference appears, since $T^{\rm E} = 0$, but $T^{\rm FP} = T - \frac{\Box}{m^2_\Psi} T$. In the Fierz-Pauli context, $\Box \Psi =0 $ needs not to be satisfied in the presence of matter, and the trace $T^{\rm FP}$ leads to the constraint $\Psi^{\rm FP} = 2 \kappa T / m^2_\Psi$. In vacuum, both cases explicitly agree.

\section{Effective energy momentum-tensors for an isothermal sphere} \label{sec:effectiveEMT}

In order to understand the implications of these different energy-momentum tensors $T_{ab}, T^{\rm E}_{ab}, T^{\rm FP}_{ab}$, we consider the simple and relevant case of an isothermal sphere \cite{BinneyBook}. The isothermal sphere is largely used as a first approximation, particularly in the context of galaxy lensing \cite{Meneghetti2021}, to model the matter content (baryonic and dark matter) of galaxies. It assumes a spherically symmetric particle distribution that satisfies the hydrostatic equilibrium [i.e. $p'(r) = - \rho(r) \Phi_{\rm N}'(r)$], where $p$ is the pressure, $\rho$ the density of matter, and $\Phi_{\rm N}$ the Newtonian potential. The equation of state is $p = \rho \sigma^2$, where $\sigma$ is the velocity dispersion, which is assumed to be constant (isothermal). This implies that 
\begin{equation} \label{eq:rhoIsothermal}
  \rho(r) = \frac{\sigma^2}{2\pi G r^2} \, . 
\end{equation}
In this context, and at rest with respect to the mean system velocity $u^i$, the energy-momentum tensor can be expressed as $(T_\mu^\nu) = \rho \mbox{ diag}(-1, \sigma^2, \sigma^2, \sigma^2)$. 

From the $T_{ab}$ as introduced above and the $T^{\rm FP}_{ab}$ definition \eqref{eq:TpsiFP}, the Fierz-Pauli effective density coincides with the isothermal one, but the pressure is different and it is no longer isotropic, that is, 
\begin{align}
  & \rho^{\rm FP} = \rho \, , \nonumber \\
  & p_r^{\rm FP} = \rho \left[ \sigma^2  + \frac{4}{3 m_\psi^2 r^2} (-1 + 3 \sigma^2)\right] \, ,\label{eq:effectiveRhoP_FP}\\ 
  & p_\theta^{\rm FP} = p_\phi^{\rm FP} =  \rho \left[ \sigma^2  - \frac{4}{3 m_\psi^2 r^2} (-1 + 3 \sigma^2)\right] \, . \nonumber
\end{align} 
In the above, it was used that, in this coordinate system denoted by $(x^\mu)$, with $x^0 = t, x^1 = r, x^2=\theta, x^3 = \phi$,
\begin{enumerate}
  \renewcommand{\labelenumi}{\roman{enumi}.}
  \setlength{\itemsep}{0pt}
  \setlength{\topsep}{0pt}
    \item $(\partial_a \partial^b - \delta^b_a\Box) T \to (\nabla_\mu \nabla^\nu - \delta^\nu_\mu \nabla^2) T$,
    \item $T = (-1 + 3 \sigma^2)\rho$,
    \item $\nabla^2 T = 2 T/r^2$,
    \item $\nabla_j \nabla^i T = \partial_j\partial^i T + \Gamma_{j r}^i T'$,
    \item and that the only non-null components of $\Gamma_{j r}^i$, at background level, are $\Gamma^\theta_{\theta r} = \Gamma^\phi_{\phi r} = 1/r$.  
\end{enumerate}

The results in Eq.~\eqref{eq:effectiveRhoP_FP} show that the effective density and pressure reduce to their standard values, $\rho$ and $p$, in the limit $m_\Psi \to \infty$. A similar behavior is observed in the limit $r \to \infty$. Since the dispersion velocity cannot exceed the speed of light, the minimal restriction is $0 < \sigma < 1$. In practice, $\sigma^2 \ll 1$, so that $T = -1 + 3\sigma^2 < 0$. Consequently, $p_r^{\rm FP} < p$ and $p_\theta^{\rm FP} > p$. It is important to note that for sufficiently small $r$, the massive corrections necessarily dominate. In this regime, according to Eq.\eqref{eq:effectiveRhoP_FP}, $p_\theta^{\rm FP}$ diverges positively, while $p_r^{\rm FP}$ tends to negative infinity.

The essential point to emphasize here is that $T^{\rm FP}_{ab}$ is manifestly distinct from $T_{ab}$ in physically relevant configurations. Such differences generally affect the absorption and lensing properties of gravitational waves. In the following, we analyze the specific case of $T^{\rm E}_{ab}$.

\bigskip

We now consider the case of the effective energy-momentum tensor $T^{\rm E}_{ab}$ \eqref{eq:def_psi_t}. First we note that, since $T$ does not depend on time, $\Box \to \nabla^2$, and that
\begin{equation}
    \nabla_i \nabla^j \nabla^{-2} T = \frac{-1+ 3 \sigma^2}{4 \pi G}   \nabla_i \nabla^j \Phi_{\rm N} \, ,
\end{equation}
since $\nabla^2 \Phi_{\rm N} = 4 \pi G \rho$. Moreover, $\nabla_r \nabla^r \Phi_{\rm N} = \Phi_{\rm N}''$, $\nabla_\theta \nabla^\theta \Phi_{\rm N} = \Phi_{\rm N}'/r = \nabla_\phi \nabla^\phi \Phi_{\rm N}$, and the other components of $\nabla_i \nabla^j \Phi_{\rm N}$ are zero. Since $\rho(r)$  describes an isothermal sphere, we have $\Phi_{\rm N} =  2 \sigma^2 \ln(r/r_0)$, where $r_0$ is a constant. Therefore, $\Phi'_{\rm N}/r = 4 \pi G \rho$, $\Phi''_{\rm N} = - 4 \pi G \rho$, and
\begin{align}
    & \nabla_r \nabla^r \nabla^{-2} T  = - (-1 + 3\sigma^2) \rho  \, ,  \\[.5cm]
    & \nabla_\theta \nabla^\theta \nabla^{-2} T  =  \nabla_\phi \nabla^\phi \nabla^{-2} T  = (-1 + 3 \sigma^2) \rho   \, . \nonumber
\end{align}
One verifies that, as expected, summing the above three components leads to $(-1 + 3 \sigma^2) \rho = T$. This is expected since $\nabla_\mu \nabla^\mu \nabla^{-2} T = \nabla^2 \nabla^{-2} T = T$.

In conclusion, in the weak field limit, the effective energy density and pressure for quadratic gravity in the extended transverse gauge are
\begin{align}
  & \rho^{\rm E} = \rho \, , \nonumber \\[.3cm]
  & p_r^{\rm E} = p - \frac{2}{3}\left(-\rho + 3 p \right) = \rho \left( \frac{2}{3} - \sigma^2\right) , \\[.3cm]
  & p_\theta^{\rm E} = p_\phi^{\rm E}  = p = \rho \sigma^2 . \nonumber
\end{align}
Since the photon gas case corresponds to $p = \rho/3$, it is safe to assume $\sigma^2 < 1/3$, implying that $\rho/3 < p^{\rm E}_r < 2\rho/3$. 

With the results above, we have illustrated how to deal with $T_{ab}^{\rm E}$ and made explicit the differences between $T_{ab}$, $T_{ab}^{\rm FP}$ and $T_{ab}^{\rm E}$ in the context of a simple but relevant physical configuration. We stress that, for the FP case, the correction $T_{ab}$ depends explicitly on the product $m_\Psi^2 r^2$, in particular, it can become arbitrarily large at small $r$; while the quadratic gravity correction to $T_{ab}$ is smooth and independent of $m_\Psi$. As remarked in the Introduction, observable consequences should be considered with care, due to the presence of the ghost. Hence, these results are more thought experiments for understanding the theory than as a direct path to perform observational tests.

\section{Comparison between Teyssandier and extended transverse gauge} \label{sec:Comparisson}

The two gauges can be made identical if the trace of the energy-momentum tensor vanishes, as in vacuum or radiation-dominated backgrounds. To achieve this identity, in the extended transverse gauge one should use $\Theta = - \Phi$ and in the Teyssandier, one should use the residual gauge to set $\Psi = \partial^a \Psi_{ab} = 0$. In this setting, Eq.~\eqref{eq:TeyssandierEqs} becomes identical to Eq.~\eqref{eq:ExtendedEqs}.

In general, the main differences are the conditions $\Psi_{ab}$ must satisfy and its corresponding induced energy-momentum tensor. Apart from that, the metric decomposition into $\tilde h_{ab}$, $\Psi_{ab}$ and $\Phi$ is also different. Regarding the induced energy-momentum tensor $T_{ab}^{\rm E}$, the appearance of $\partial_a \partial_b / \Box^{-1}$ inside it, in general, may lead to computational difficulties. But, as we have explicitly illustrated, it is possible to solve the difficulties by introducing a potential (for the static case example, as detailed here, the Newtonian potential was sufficient). As we have detailed here, the structure of $T^{\rm E}_{ab}$ is quite similar to $T^{\rm FP}_{ab}$, which opens possibilities of comparison. However, $T^{\rm FP}_{ab}$ depends directly on $m^2_\Psi$, which appears in the same place of the operator $\Box^{-1}$ in $T^{\rm E}_{ab}$. Both $T^{\rm FP}_{ab}$ and $T^{\rm E}_{ab}$ are divergence-free tensors, contrary to $T^{\rm T}_{ab}$. Among them, only $T^{\rm E}_{ab}$ is traceless, which is necessary since $\Psi = 0$ in this gauge.

At last, considering the starting principles of each gauge, while the Teyssandier gauge is based on the gauge freedom of the vector $\Gamma_a$, the extended transverse comes straight from the Lie derivative of the metric perturbations.

\section{Conclusions} \label{sec:conclusions}

We investigated quadratic gravity, focusing on the structure of perturbative modes and the crucial role of gauge choices in simplifying the analysis of higher-derivative field equations. The main goal was to present a detailed and systematic study of the generalized transverse gauge condition, along with its residual gauge symmetries. This gauge provides a novel and effective framework for identifying and analyzing new solutions within the theory, providing clearer insight into the underlying physical degrees of freedom.

In the generalized transverse gauge, we demonstrate that the metric perturbation can be decomposed into three distinct components: a massless spin-2 field, a massive spin-2 mode, and a scalar field. This structure is qualitatively similar to the decomposition in the Teyssandier gauge, although the specific forms differ, as shown in Eqs.~\eqref{eq:TeyssandierDecomp} and \eqref{eq:ExtendedDecomp}. Importantly, this decomposition reveals that the massive spin-2 field couples to an effective energy-momentum tensor that is divergence-free and resembles the Fierz-Pauli tensor, in contrast to the Teyssandier case. The explicit forms of these energy-momentum tensors are given in Eqs.~\eqref{eq:TeyssandierT}, \eqref{eq:def_psi_t}, and \eqref{eq:TpsiFP}.

We understand that, by showing the role of gauge choices, particularly highlighting the properties of the generalized transverse gauge, and stressing the differences with respect to the Fierz-Pauli massive spin-2 mode, this work improves the tools available for studying and extending higher-derivative gravity. 

\section*{Acknowledgements}
MFSA thanks \textit{Fundação de Amparo à Pesquisa e Inovação do Espírito Santo} (FAPES, Brazil) for support. LGM thanks \textit{Conselho Nacional de Desenvolvimento Científico e Tecnológico} (CNPq, Brazil) for partial financial support—Grant: 307901/2022-0 (LGM). DCR thanks \textit{Centro Brasileiro de Pesquisas Físicas} (CBPF) and \textit{Núcleo de Informação C\&T e Biblioteca} (NIB/CBPF) for hospitality, where part of this work was done. He also acknowledges CNPq (Brazil), FAPES (Brazil) and \textit{Fundação de Apoio ao Desenvolvimento da Computação Científica} (FACC, Brazil) for partial support. 

\appendix

\section{Extended transverse gauge with arbitrary $\Theta$} \label{sec:app}

The choice $\Theta = -\Phi$ is a special one as detailed in this work. Here we consider further developments with a general $\Theta$. From Eq.~\eqref{eq:boxPsi_general}, for arbitrary $\Theta$,
\begin{equation}
    \Theta + \Phi = - \frac{1}{6} (\Psi + H)\, ,
\end{equation}
where $H$ is a harmonic function. Since $\Psi$ has no harmonic part, $H/6$ is the harmonic part of $\Theta + \Phi$, implying $H=0$ for $\Theta = -\Phi$.

Using the above in Eq.~\eqref{eq:boxPsi_general0},
\begin{gather}
-\frac{1}{2 \alpha}\Box \Psi_{ab}+\frac{1}{2}\Box ^{2}(\tilde{h}_{ab}+\Psi
_{ab})  \notag \\
+(\partial _{a}\partial _{b}-\eta _{ab}\Box )\left( - \frac{\Psi + H}{6 \alpha} +\frac{1}{6}\Box (\tilde{h}+\Psi )\right) =0\,,
\end{gather}
hence,
\begin{gather}
\left( \Box - \frac{1}{\alpha}\right) \left[\Box \Psi_{ab}  + \frac{1}{3}(\partial_{a}\partial _{b}-\eta _{ab}\Box ) \Psi \right] - \nonumber  \\ 
- \frac{1}{3 \alpha} \partial_a \partial_b H =  
2\kappa \left[\Box T_{ab}  
+\frac{1}{3}(\partial_{a}\partial _{b}-\eta _{ab}\Box )   T \right]\,. \label{eq:AppGeneralPsi}
\end{gather}
This is the general $\Psi_{ab}$ equation for any $\Theta$. 

If $H = 0$, the above equation has a particular solution given by
\begin{equation} \label{eq:AppParticularPsiSol}
    \left( \Box - \frac{1}{\alpha}\right) \Psi_{ab} = 2 \kappa T_{ab} \, .
\end{equation}
For $\Theta = - \Phi$, since $\Psi = 0$, one sees that the validity of the above equation is limited, since it only holds for $T = 0$. In this limit, indeed $T^{\rm E}_{ab} = - T_{ab}$, and therefore there is no contradiction with the previous results.

From the trace of Eq.~\eqref{eq:AppParticularPsiSol}, and the scalar mode equation \eqref{eq:ETscalar},
\begin{equation}
    -\frac{1}{6} \left(\Box - \frac{1}{\alpha} \right) \Psi = \left(\Box - \frac{1}{3 \gamma} \right)\Phi \, .
\end{equation}
Using Eq.~\eqref{eq:boxPsi_general},
\begin{equation}
    \left(\Box - \frac{1}{\alpha} \right) \Theta = \left (\frac{1}{\alpha} - \frac{1}{3 \gamma} \right) \Phi \, .
\end{equation}

The above equation is satisfied immediately for $\Theta = -\Phi$, without further restrictions (apart from $T=0$). For any choice of the form $\Theta = k \Phi$, with $k\not=-1$, the single solution (apart from $\Phi=0$) is $\alpha = 3 \gamma$, that is, $m_\Phi^2 = m_\Psi^2$. This particular case was explored in some works before (e.g., \cite{Teyssandier1989}).

It is interesting to point out the existence of the general $\Psi_{ab}$ equation \eqref{eq:AppGeneralPsi} that is valid for any $\Theta$ and that it is independent of $\Phi$ (at least apart from a harmonic component). Although this is not a simple equation to solve, it may prove useful for general properties. In addition, $\Theta$ choices different from $\Theta = -\Phi$ can be useful for developments in particular scenarios.


%

\end{document}